\documentclass[%
 aip,
 amsmath,amssymb,
 reprint,%
]{revtex4-1}

\usepackage{graphicx}
\usepackage{dcolumn}
\usepackage{bm}
\usepackage[utf8]{inputenc}
\usepackage[T1]{fontenc}
\usepackage{mathptmx}
\usepackage{etoolbox}
\usepackage{todonotes}
\usepackage{amsmath}

\makeatletter
\def\@email#1#2{%
 \endgroup
 \patchcmd{\titleblock@produce}
  {\frontmatter@RRAPformat}
  {\frontmatter@RRAPformat{\produce@RRAP{*#1\href{mailto:#2}{#2}}}\frontmatter@RRAPformat}
  {}{}
}%
\makeatother

\begin{document}

\title{Multi-Diagnostic Characterization of Laser-Produced Tin Plasmas for EUV Lithography} 

\author{S. Musikhin}
\email{smusikhin@pppl.gov}
\affiliation{Princeton Plasma Physics Laboratory, Princeton, NJ, 08540 USA}
\author{A. Morozov}
\affiliation{Department of Mechanical and Aerospace Engineering, Princeton University, Princeton, NJ 08540, USA}
\author{A. Griffith}
\author{S. Yatom}
\author{A. Diallo}
\affiliation{Princeton Plasma Physics Laboratory, Princeton, NJ, 08540 USA}
\date{\today}

\begin{abstract}
We present a comprehensive characterization of laser-produced tin (Sn) plasmas relevant to extreme ultraviolet (EUV) lithography using a multi-diagnostic suite integrated into the new experimental platform, "SparkLight". Tin plasmas are generated by irradiating a continuously moving tin-coated wire with laser pulses (1064 nm, 10 ns, up to $5.7\times10^{10}$ W/cm$^2$) and probed via coherent Thomson scattering, laser interferometry, and EUV emission spectroscopy. Thomson scattering measurements reveal electron temperatures and densities that decay with distance from the target. Densities derived from Thomson scattering are cross-validated against laser interferometry, showing excellent agreement. Correlating the results of these laser diagnostics with spatially resolved EUV spectroscopy suggests that the bulk of useful EUV emission originates within 150 $\mu$m of the target and is generated under suboptimal plasma conditions. This work demonstrates a practical integrated approach for plasma characterization in EUV source development.
\end{abstract}

\maketitle 

\section{Introduction}
Extreme ultraviolet (EUV) lithography at 13.5 nm has become the enabling technology for the "sub-7 nm" technology node in semiconductor manufacturing, with laser-produced tin (Sn) plasmas serving as the primary photon source. The transition from deep ultraviolet (193 nm) to EUV wavelengths has enabled continued feature-size reduction and density scaling, driving the demand for increasingly efficient and well-characterized EUV photon sources. 

A critical challenge in EUV source development is maximizing its spectral purity and conversion efficiency.\cite{bakshi2023} Spectral purity represents the emission power within the 2\% bandwidth at 13.5 nm, the "in-band" region required by Mo/Si mirror optics and resist chemistry, relative to the full-spectrum power. Conversion efficiency is the ratio of in-band 2$\pi$ emission to incident drive-laser pulse power. Optimizing conversion efficiency and spectral purity requires precise control of interdependent plasma parameters: electron density ($n_\text{e}$), electron temperature ($T_\text{e}$), and tin ion charge state distribution. These quantities determine the EUV emissivity and the optical depth of the plasma at 13.5 nm. \cite{bakshi2023} For example, high-efficiency in-band emission was shown to be achievable at plasma temperatures between 25\textendash50 eV. \cite{nishihara2008a, white2005, poirier2006} However, because tin plasma is optically thick to 13.5 nm light, the final EUV output is also determined by the opacity, which is a function of electron density. \cite{fujioka2005, pan2023, tallents2019, versolato2019} Finally, controlling the tin ion charge state distribution is essential to maximize spectral purity. Accurately measuring these plasma parameters is also critical for validation of the radiation hydrodynamic codes that are used in optimization of EUV sources. 

We present ''SparkLight", a new experimental platform for generating and comprehensively characterizing laser-produced tin plasmas. Tin plasmas are generated by irradiating a continuously moving tin-coated wire with a 1064 nm Nd:YAG laser. The facility integrates three complementary diagnostics: EUV emission spectroscopy, laser interferometry, and coherent Thomson scattering, enabling time- and space-resolved measurements of $n_\text{e}$ and $T_\text{e}$. 

We devote particular attention to Thomson scattering, which is a critical diagnostic for non-perturbative, space- and time-resolved characterization of laser-produced plasmas. When the scattering signal is separated by polarization, the polarized Thomson scattering signal can be isolated from the intense unpolarized plasma self-emission. \cite{bak2025, swadling2022, kieft2005, sato2019a} Previous Thomson scattering implementations on tin plasmas have relied on triple- or six-grating spectrometers to reject intense plasma self-emission.\cite{sato2019a, kieft2005} Our approach achieves comparable performance using a compact, cost-effective design based on a Wollaston prism, \cite{swadling2022} two volume Bragg grating (VBG) notch filters,\cite{klarenaar2015, yatom2022, bak2023} and a single-grating spectrometer, substantially lowering the barrier to the adoption of this diagnostic technique.

\section{Experimental}
\subsection{Target System and Drive Laser Architecture}

\begin{figure*}
    \centering
    \includegraphics[width=1\textwidth]{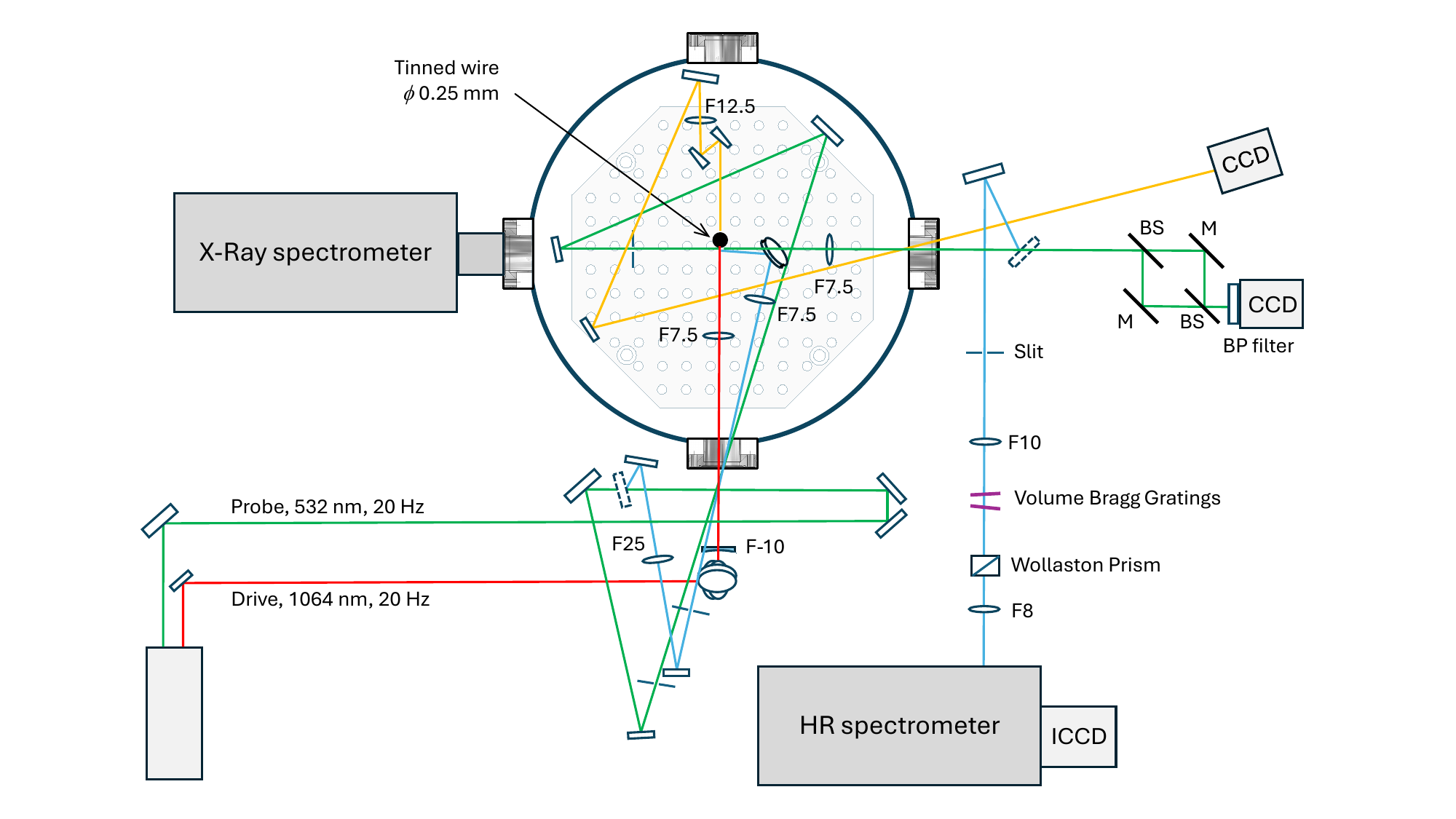}
    \caption{Schematic of the "SparkLight" facility showing a vacuum chamber, target (tinned wire), Thomson scattering, laser interferometry, and EUV emission spectroscopy diagnostics. Optical paths are color-coded: red\textemdash drive beam, green\textemdash 532 nm probe beam used for laser interferometry and Thomson scattering (part of the optical path related to Thomson scattering shown in blue), and yellow\textemdash laser beam profiling. Acronyms are defined as follows: F: optical focal length in cm, M: mirror,  BS: beamsplitter, BP: bandpass.}
    \label{whole-setup}
\end{figure*}

Figure \ref{whole-setup} shows the overall experimental setup of the "SparkLight" facility with a vacuum chamber, target, and optical diagnostics. The base chamber pressure was 10$^{-4}$ mTorr, rising to $\approx$ 1 mTorr during experiments due to continuous wire ablation. Tin plasmas were generated by irradiating a 250 $\mu$m diameter tin-coated copper wire (Goodfellow Inc., tin coating thickness $\geq0.3 \mu m$) with a 1064 nm Nd:YAG laser (Quantel Ultra, 10 ns FWHM,  20 Hz). The laser beam was focused on the wire with spot sizes, $D$, between 75\textendash100 $\mu$m (FWHM) while the laser energies delivered to the target varied between 12\textendash38 mJ. We calculated peak laser intensities for a Gaussian profile as $I=( 2 \sqrt{\frac{2 \ln 2}{2 \pi}} )^3 \frac{E_\text{laser}}{\tau D^2}$, where $E_\text{laser}$ is the pulse laser energy and $\tau$ is the FWHM of the laser temporal profile. The spot size and intensity distribution in the focal plane was monitored with a CCD camera (Thorlabs, DCU223) by relaying the laser beam through a doublet achromat lens after Fresnel reflections from two optical wedges (yellow line in Figure \ref{whole-setup}).

To ensure that each laser pulse hits a "fresh" surface of the wire, it was continuously pulled by a motor at a speed of $\approx$10 mm/s. We estimated the transverse displacement of the wire during pulling relative to the drive beam axis at under 30 $\mu$m. The assembly was installed inside a vacuum chamber on a high precision vacuum compatible x,y,z-stage, which was controlled remotely to adjust the wire position.

The resulting extreme ultraviolet (EUV) emission was detected with a high-resolution X-ray spectrometer (hpSpectroscopy GmbH, highLIGHT, 100 $\mu$m horizontal entrance slit) coupled with a CCD camera (hpSpectroscopy GmbH, maxCAM, 1024$\times$256 pix, 26 $\mu$m/pix). The achieved spectral resolution was 0.05 nm, limited by the CCD pixel size and the slit width. The signal was accumulated over 10 s, which corresponds to 200 laser shots. The spectrometer was calibrated using identified Li II, Li III, F V, F VI, and F VII lines observed in laser-produced plasma from a LiF solid target.

\subsection{Polarization-Separated Thomson Scattering Diagnostic}
\begin{figure} [htbp]
    \centering
    \includegraphics[width=1\linewidth]{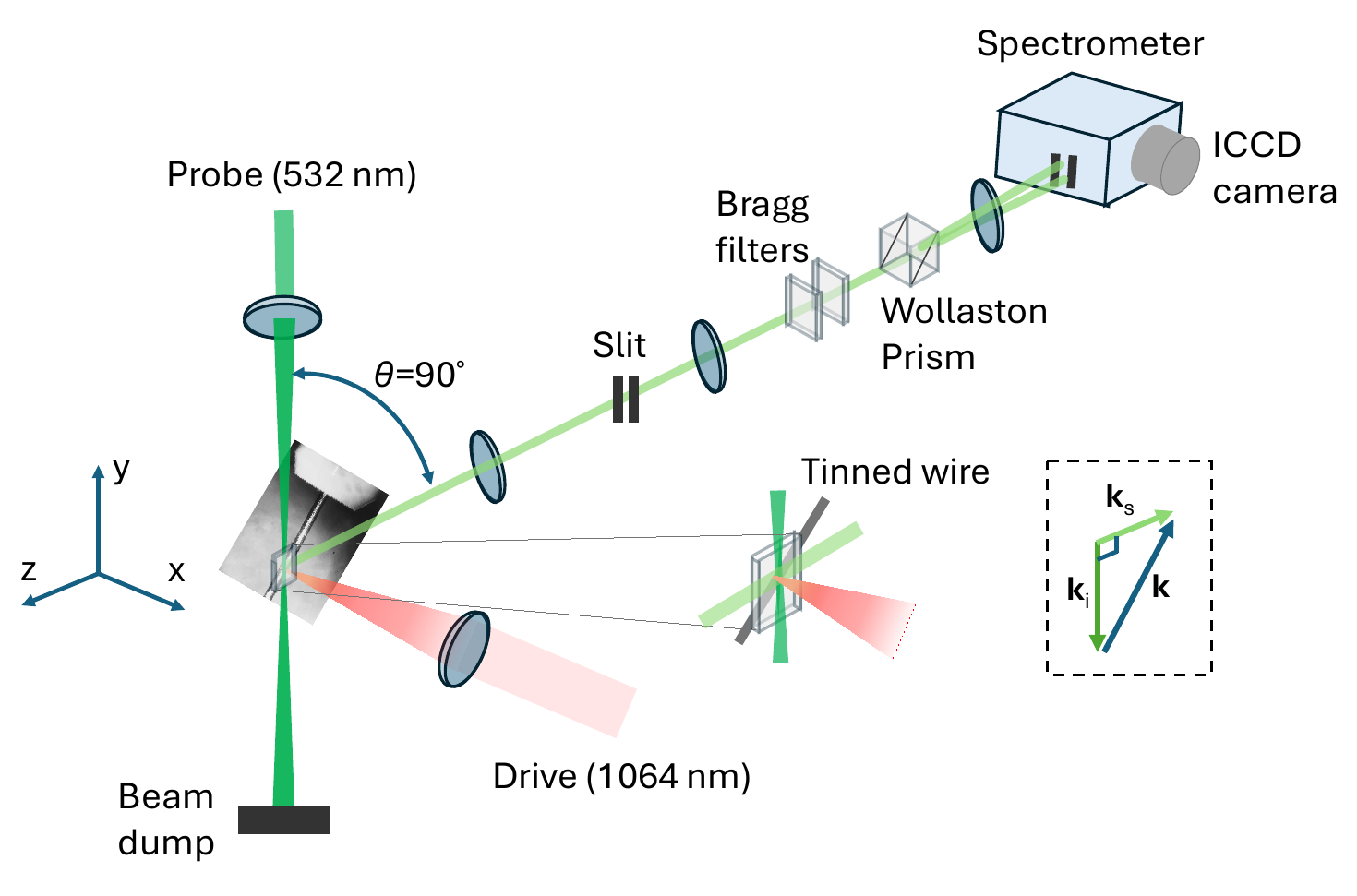}
    \caption{Schematic of the compact polarization-separated Thomson scattering diagnostic. A Wollaston prism splits the collected signal into s- and p-polarized components.}
    \label{ThomsonSetup}
\end{figure}

Figure \ref{ThomsonSetup} shows the schematic of the compact polarization-separated Thomson scattering diagnostic. The probe beam was oriented at 90 degrees to the drive beam. The incident light wave vector is denoted as $\textbf{k}_\text{i}$. The scattering signal was collected at 90 degrees to the probe laser with an achromat lens, providing an intermediate image of the plasma on a narrow slit, which served to reduce contributions from stray light and the plasma self-emission. Then, the spatially filtered image was relayed to the spectrometer entrance slit with two achromat lenses. The scattered light wave vector is denoted as $\textbf{k}_\text{s}$. To spatially characterize the plasma with respect to the target, the tinned wire was scanned along the $x$-direction as defined in Figure~\ref{ThomsonSetup}. 

The probe beam was derived from the second harmonic of the drive laser (532 nm, 10 ns FWHM, 20 mJ/pulse) and focused to a 40 $\mu$m FWHM spot at the plasma. Therefore, a single laser generated both drive and probe pulses. The optical paths for the drive and probe pulses were adjusted using an optical delay line so that the probe pulse reached the plasma volume near the peak of the drive pulse. This configuration eliminated jitter and ensured temporal synchronization between plasma generation and probing. 

Stray laser light was effectively suppressed using two volume Bragg grating (VBGs, OptiGrate, FWHM spectral bandwidth <10 cm$^{-1}$) notch filters. VBGs provide ultra-narrow notch filtering with theoretical attenuation of OD 3\textendash4 per element, while maintaining >95\% transmittance outside the notch. Both VBGs were placed in the collimated part of the scattered beam to maximize their performance and achieve high attenuation around the 532 nm laser line.

The collected light was divided into two orthogonally polarized beams using a quartz Wollaston prism (Thorlabs, specified extinction ratio $10^4:1$). Here, the signal polarization was defined relative to the scattering plane, which is the plane containing wave vectors $\textbf{k}_\text{i}$ and $\textbf{k}_\text{s}$, as shown in Figure \ref{ThomsonSetup}. Hence, the s-polarized component was perpendicular to the scattering plane (the same as the probe beam), and the p-polarized component was parallel to the scattering plane. Both beams were focused on the entrance slit of the imaging spectrometer (Acton Research SpectraPro 300i, 1200 and 2400 g/mm) with a small vertical separation. The entrance slit was set to 125 $\mu$m to match the probe beam waist determined from Rayleigh scattering images in atmospheric air. We determined spectral resolution as the FWHM of the Rayleigh scattering signal, measuring 0.3 nm with the 1200 g/mm grating and 0.12 nm with the 2400 g/mm grating.

Finally, the signals were imaged with a gated emICCD camera (PI MAX4, 10 ns gate). For each distance from the target, ten images were captured, while each image was an accumulation of 20–40 laser shots. 

\subsection{Laser Interferometry Diagnostic}
For laser interferometry, we used a small fraction of the collimated 532 nm beam ($\varnothing$ 6 mm) directed through the plasma (Figure \ref{whole-setup}). The plasma image was formed with an achromatic doublet (f = 75 mm, $\varnothing$ 1”) on the 8-bit CCD camera (Thorlabs DCU223, 4.65 $\mu$m/pix). Before reaching the CCD, the probe beam was split into signal and reference beams and then recombined using two beamsplitters in a compact folded Mach-Zehnder interferometer positioned after the imaging lens. Part of the $\varnothing$ 6 mm beam propagated through the plasma (signal), imprinting phase information onto its wavefront, while part of the beam bypassed the plasma (reference). The accumulated phase change in the plasma resulted in a fringe shift in the recorded interferogram. To filter out the background visible radiation from plasma we used a bandpass filter (Thorlabs FLH532-10). The time resolution in the measurements was defined by the probe pulse duration of 10 ns while the attained spatial resolution was about 5 $\mu$m.

The interferograms were converted to phase shift maps using IDEA software \cite{hipp2004}. The 2D electron density maps were then reconstructed by inverse Abel transform. We assumed that the plasma is cylindrically symmetric; given that the 100 $\mu$m laser spot was centered on the 250 $\mu$m wire, we approximated the target as a planar surface and neglected the curvature of the wire. The Abel transform was performed using the basis set expansion algorithm ("basex") \cite{dribinski2002} implemented in the pyAbel Python package. \cite{hickstein2019}

\section{Results and Discussion}
\subsection{Plasma Emission Spectroscopy}
The EUV emission of tin plasma is primarily characterized by the Unresolved Transition Array (UTA), which is a quasi-continuum emission of many closely packed transitions of highly charged ions. The peak of the UTA emission is known to shift toward shorter wavelengths as the ionization state increases. \cite{versolato2019, schupp2019a} Specifically, Sn$^{10+}$ emits near 14 nm, Sn$^{12+}$ near 13.5 nm, and Sn$^{14+}$  near 13 nm.\cite{churilov2006, churilov2006a, ryabtsev2008, versolato2019} 

We assume the plasmas investigated here consist primarily of tin ions with minimal contamination from the copper. To ensure this, we minimize the effect of laser pulse ablating through the tin coating by continuously pulling the wire during the experiment so that each laser pulse hits a previously unexposed surface of the coating. Separate tests with multiple laser exposures of the same spot yielded a distinct copper emission spectrum, which was not observed during our single-exposure experiments. However, due to spectral overlap of tin and copper features, we cannot rule out a minor copper contamination in our plasmas.

\begin{figure}
    \centering
    \includegraphics[width=\linewidth]{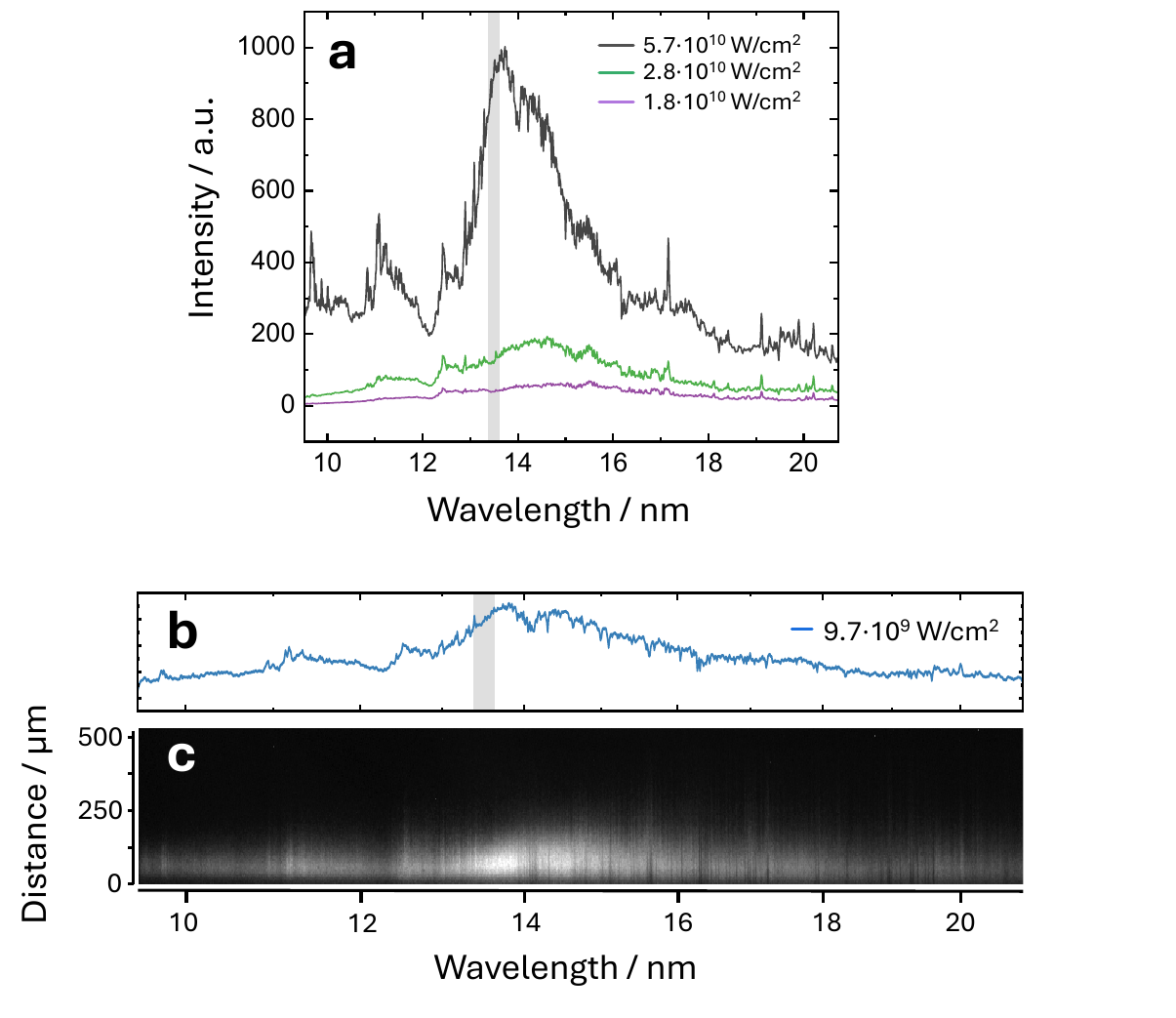}
    \caption{(a) Emission spectra from a laser-produced tin plasma obtained at different laser intensities. (b) and (c) Spatially and spectrally resolved emission spectrum obtained at the laser intensity value used during Thomson scattering experiments. All spectra were accumulated over 200 laser shots (1064 nm, 10 ns) onto a tin-coated wire.}
    \label{EUVspectrum}
\end{figure}

Figure \ref{EUVspectrum}a shows the EUV emission spectra obtained at different laser intensities. The in-band region is shown in shaded gray. The most prominent feature is situated between 13\textendash15 nm, which is attributed to multiple resonant transitions in Sn$^{8+}$.. Sn$^{14+}$.\cite{churilov2006, churilov2006a, churilov2006b, ryabtsev2008, schupp2019a} Other bands observed in the spectra, between 10\textendash12 nm and 16\textendash20 nm, correspond to Sn$^{5+}$.. Sn$^{9+}$ ions. \cite{churilov2006, churilov2006b, torretti2018, versolato2019, schupp2019a} 

The peak emission feature blue-shifts with increasing laser intensity, which is in line with the existing observations, and explained by the creation of tin ions of higher charge state.\cite{schupp2019a, versolato2019} At the highest drive laser intensity used here, $5.7\times10^{10}$ W/cm$^2$, the emission peaks at 13.6\textendash13.8 nm, overlapping only partially with the 13.5$\pm$0.135 nm in-band region. This red-shift with respect to the in-band region suggests that the plasma conditions, particularly $T_\text{e}$, were suboptimal for maximum in-band emission. Experiments deploying Nd:YAG lasers showed that more optimal electron temperatures of 25\textendash50 eV can be achieved at laser intensities on the order of 10$^{11}$ W/cm$^2$.\cite{schupp2019a} 

However, spectral data alone are insufficient to definitively determine the plasma temperature without comprehensive radiation-hydrodynamic simulations due to the emission self absorption and complexity of transitions from multiple Sn charge states. \cite{versolato2019} For example, Pan et al. reported that spectral red-shifts in laser-produced tin plasmas can occur even when the electron temperature remains constant.\cite{pan2023} By modeling radiative transfer, the authors attributed the red-shift to a reduced optical depth as $n_\text{e}$ decreased with distance, rather than to a change in $T_\text{e}$. Therefore, specialized diagnostics such as Thomson scattering are required.

Figures \ref{EUVspectrum}b and c show the spectrally and spatially resolved emission spectrum obtained at the laser intensity of $9.7\times10^{9}$ W/cm$^2$. The laser intensity was limited to this value during Thomson scattering experiments because the energy from a single laser was split to generate both drive and probe laser pulses (see Figure \ref{whole-setup}). To obtain a spatially resolved EUV plasma image, we placed an additional 40 $\mu$m vertical slit in front of the spectrometer entrance. Figure \ref{EUVspectrum}c shows the resulting obscure image of the plasma on the CCD, which indicates that the bulk of the EUV emission originated within 150 $\mu$m of the ablated surface.

\subsection{Thomson Scattering}
\begin{figure}
    \centering
    \includegraphics[width=1\linewidth]{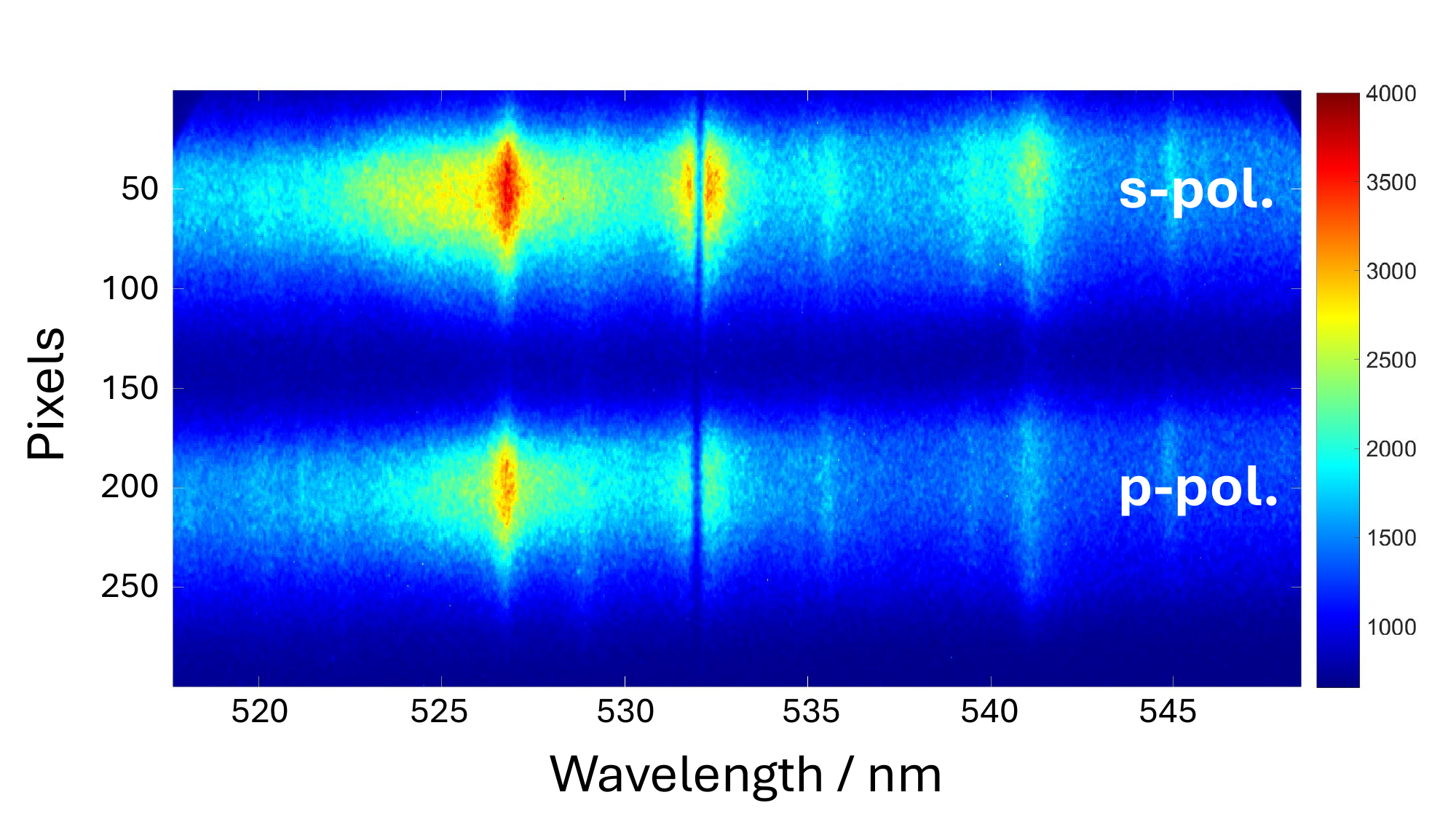}
    \caption{Signals separated by polarization using a Wollaston prism and imaged with an emICCD camera. The s-polarized signal consists of Thomson scattering and the s-polarized part of the plasma self-emission, the p-polarized signal contains the remaining part of the unpolarized self-emission. The signals were collected using a 1200 g/mm grating, $270\pm20\ \mu m$ from the target, integrated over 40 laser shots per sequence, and averaged over 10 sequences.}
    \label{highlowband-example}
\end{figure}

For Thomson scattering diagnostics, the collected light was split into s- and p-polarized components using a Wollaston prism. Separation of signals by polarization with subsequent subtraction of the unpolarized background emission has been utilized in plasma scattering diagnostics before. \cite{bak2025, swadling2022, kieft2005, sato2019a} Here, the probe laser was s-polarized relative to the scattering plane. Since Thomson scattering preserves the probe polarization, the s-polarized signal consisted of Thomson scattering and half of the unpolarized plasma self-emission, while the p-polarized signal contained the remaining half of the unpolarized self-emission (Figure \ref{highlowband-example}). Thus, the self-emission contribution could be subtracted to isolate the Thomson scattering signal. Before subtracting, the signals were corrected for the transmission efficiency of the system for different polarizations. Here, the effects of light depolarization via relativistic effects and Faraday rotation were negligible, as these only become appreciable at electron temperatures $T_\text{e} \geq10$ keV and laser intensities $I \geq10^{12}$ W/cm$^2$, respectively. \cite{sheffield2011} 

The signal shown in Figure \ref{highlowband-example} contains clear line emission features, as it was collected at a further distance ($270\pm20\ \mu m$) from the target, where the line emission exceeds the continuum self-emission.\cite{harilal2005, harilal2022} Signals collected closer to the target, e.g., at distances of 120\textendash150 $\mu$m, were dominated by continuum emission. 

\begin{figure}
    \centering
    \includegraphics[width=1\linewidth]{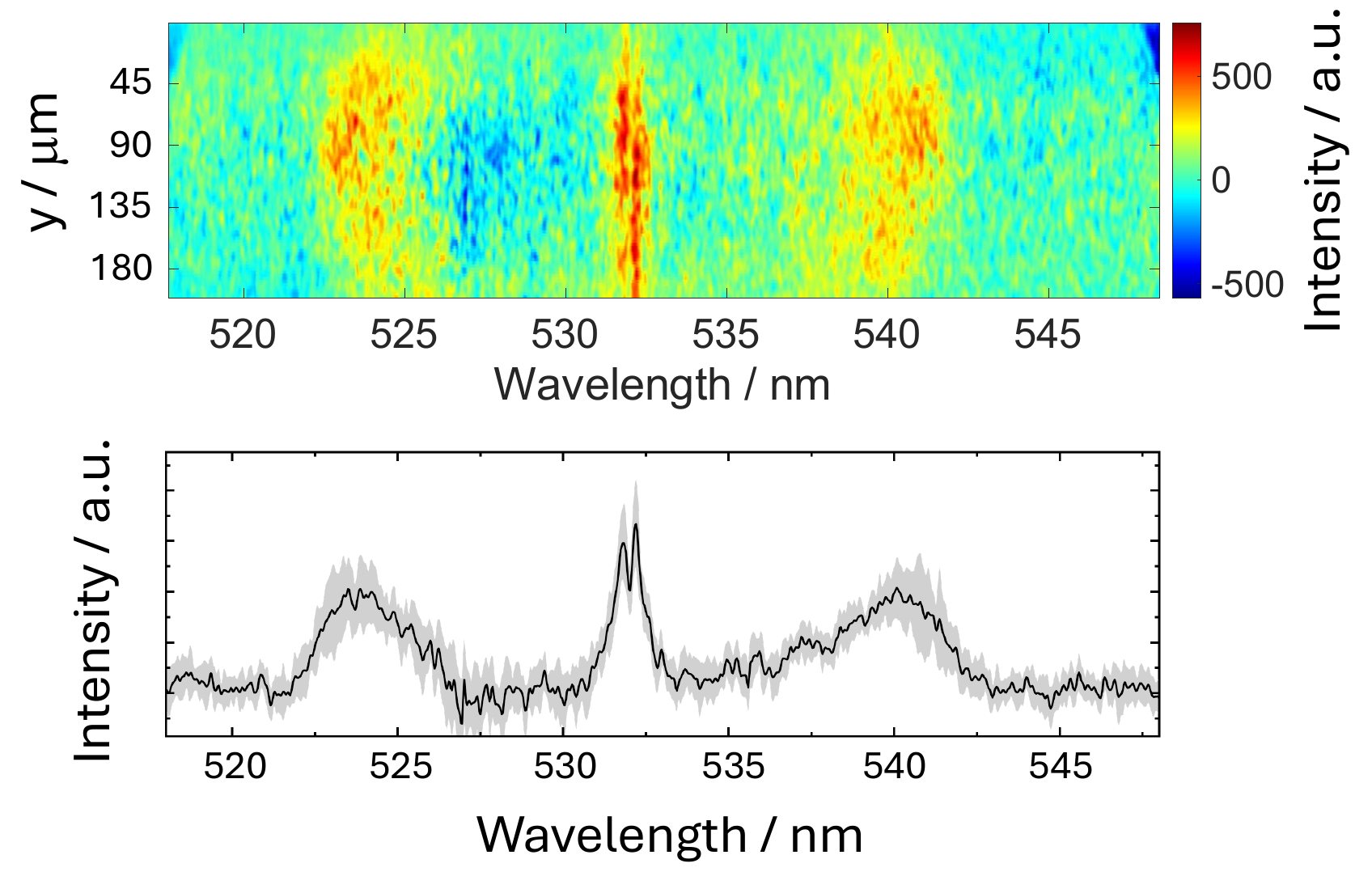}
    \caption{(Top) 2D image of the Thomson scattering signal. (Bottom) Vertically integrated spectral intensity showing distinct sidebands (electron plasma features, EPW) and a strong central feature attributed to an under-resolved ion acoustic wave (IAW). The signal was collected using a 1200 g/mm grating, $270\pm20\ \mu m$, integrated over 40 laser shots per sequence, and averaged over 10 sequences.}
    \label{Thomsonexample}
\end{figure}

Figure \ref{Thomsonexample} shows an example of the Thomson scattering spectrum. The signal was collected at a distance of $270\pm20\ \mu m$ from the target, integrated over 40 laser shots per sequence, and averaged over 10 sequences. The top figure shows the 2D spectral image after subtracting the plasma self-emission, while the bottom figure demonstrates the corresponding vertically integrated spectral intensity. The spectrum includes a coherent (collective) electron plasma feature (EPW) and a strong central feature, which we attributed to an under-resolved ion acoustic wave (IAW) scattering.

The degree of coherent effects in the Thomson signal is given by the scattering parameter, $\alpha=\frac{1}{k\lambda_\text{D}}=\frac{\lambda_\text{laser}}{4\pi \sin({\theta/2})\lambda_\text{D}}$, where $k$ is the wavenumber of the incident light, $\lambda_\text{D}$ is the Debye length, $\lambda_\text{laser}$ is the laser wavelength, and $\theta$ is the scattering angle. For the plasma conditions investigated here, $\alpha=1.7 \text{--} 3.2$, i.e., the scattering signal was dominated by coherent (collective) plasma oscillations.  

\begin{figure} 
    \centering
    \includegraphics[width=0.8\linewidth]{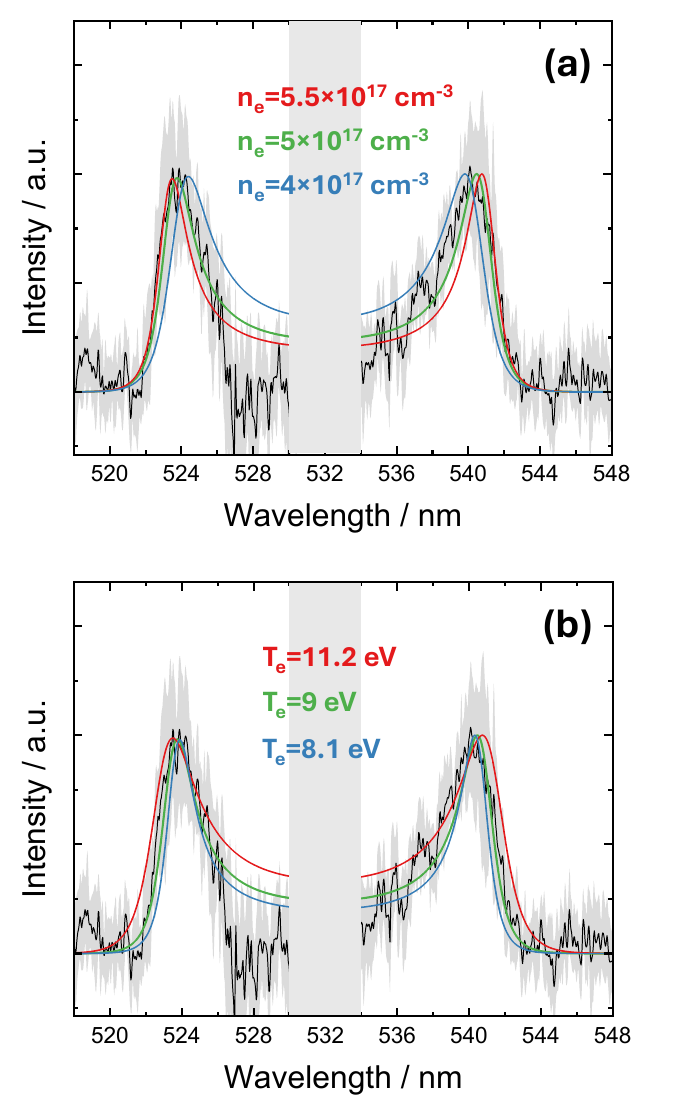}
    \caption{Normalized Thomson scattering spectrum and theoretical fits to the coherent scattering feature. The green line shows the best fit achieved at $T_\text{e} = 9$ eV and $n_\text{e} = 5 \times 10^{17}$ cm$^{-3}$, while blue and red lines demonstrate the fit uncertainty found by varying $n_\text{e}$ in plot (a), and $T_\text{e}$ in plot (b). The central region, $532\pm2$ nm (shaded gray), was excluded from the fit. The signal was collected using a 1200 g/mm grating, $270\pm20\ \mu m$ from the target, integrated over 40 laser shots per sequence, and averaged over 10 sequences. The inferred plasma parameters were $T_\text{e} = 9^{+2.25}_{-0.9}$ eV and $n_\text{e} = 5^{+0.5}_{-1} \times 10^{17}$ cm$^{-3}$.}
    \label{Thomson-collective-fit}
\end{figure}

To extract plasma parameters, we fit the experimental spectra with a full theoretical Thomson scattering model, as shown in Figure \ref{Thomson-collective-fit}. We used a spectral density function, $S(\mathbf{k},\omega)$, derived by Sheffield \cite{sheffield2011} and implemented in the PlasmaPy Python package \cite{PlasmaPyCommunity2025}. The model assumes a Maxwellian plasma, which is justified considering the rapid electron thermalization relative to the diagnostic timescale. For typical EUV laser-produced plasmas ($n_\text{e}=10^{18}\text{--}10^{19}$ cm$^{-3}$ and $T_\text{e}=10\text{--}50$ eV), the $e\text{--}e$ collision frequency, $\nu_\text{ee} \approx 2.91 \times 10^{-6} \frac{n_\text{e} \ln{\Lambda}}{T_\text{e}^{3/2}}$ [s$^{-1}$], \cite{nrl2023} where $\ln{\Lambda}$ is the Coulomb logarithm, $n_\text{e}$ is in cm$^{-3}$, and $T_\text{e}$ is in eV, corresponds to collision times on the order of picoseconds, while the timescale of a Thomson scattering measurement is 10 nanoseconds. Model parameters, $T_\text{e}$ and $n_\text{e}$, were optimized via non-linear least-squares minimization using a differential-evolution algorithm. Before fitting, theoretical spectra were convolved with the instrument function, modeled as a Gaussian kernel with FWHM = 0.3 nm (spectral resolution). The central region, $532\pm2$ nm (shaded gray), which includes an IAW, was excluded from the fit. Between 527\textendash529 nm, the signal falls below zero intensity, which is an artifact of the background subtraction procedure. We determined the uncertainty by varying $n_\text{e}$ (Figure \ref{Thomson-collective-fit}a) and $T_\text{e}$ (Figure \ref{Thomson-collective-fit}b) until the fit fell outside of the noise of the experimental spectrum. For the experimental signal shown in Figure \ref{Thomson-collective-fit}, the inferred plasma parameters were $T_\text{e} = 9^{+2.25}_{-0.9}$ eV and $n_\text{e} = 5^{+0.5}_{-1} \times 10^{17}$ cm$^{-3}$.

\begin{figure}
    \centering
    \includegraphics[width=0.9\linewidth]{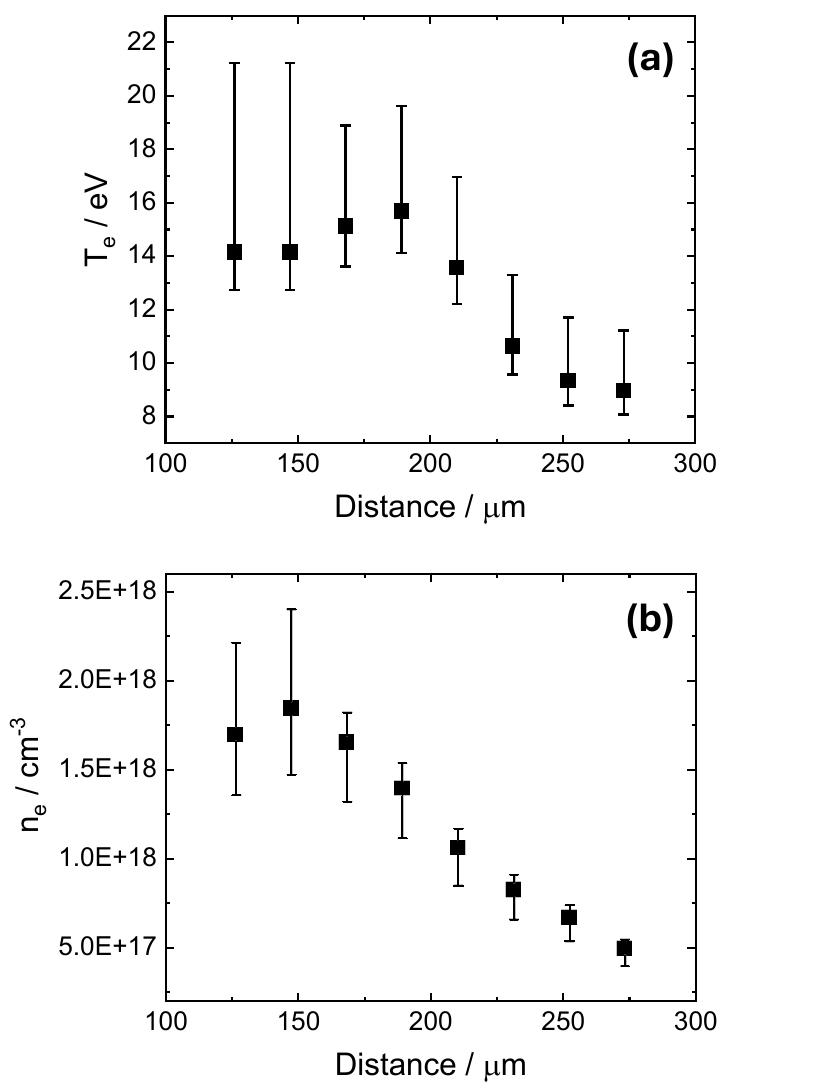}
    \caption{Spatial profiles of the (a) electron temperature ($T_\text{e}$) and (b) electron density ($n_\text{e}$) obtained by fitting the coherent component of Thomson scattering.}
    \label{boxplot_coherent}
\end{figure}

To obtain the spatial evolution of $T_\text{e}$ and $n_\text{e}$ (Figure \ref{boxplot_coherent}), we analyzed the Thomson signals collected at different distances from the target. The error bars correspond to fit uncertainties, as described above. We collected measurable signals at distances between 120\textendash270 $\mu$m from the target. Measurements further from the target were deemed irrelevant, as the plasma temperature is insufficient for EUV emission. Measurements closer to the target, where $n_\text{e}$ and $T_\text{e}$ are the highest, were unsuccessful due to dominant Bremsstrahlung emission in dense plasmas and spectral broadening of EPW features that exceeded the detector range. Collecting signals from that region would require higher probe pulse energy and either resolving only one of the two EPW features or increasing the detection spectral range to resolve both. This remains the focus of a future investigation.

The maximum measured electron temperature ( $T_\text{e}=15$ eV,) falls significantly below the 25\textendash50 eV optimal range for in-band EUV emission.\cite{nishihara2008a, white2005, poirier2006} This is consistent with the observed EUV emission spectrum, which was red-shifted relative to the in-band region (Figure \ref{EUVspectrum}b). 

We note that we cannot confidently claim that the electron temperatures plateau around 15 eV, as may appear in Figure \ref{boxplot_coherent}. Firstly, there is a high degree of uncertainty associated with the data between 120\textendash170 $\mu $m from the target, as these measurements were impacted by the limited spectral range of the grating and low signal-to-noise ratios, as explained above. Secondly, Figure \ref{EUVspectrum}c shows that the bulk of the EUV emission originates within 150 $\mu $m from the target, implying that $T_\text{e}$ is higher in this region compared to the larger distances. However, in our current configuration, this near-target region remains inaccessible to the Thomson scattering diagnostic. Excluding data from the 120\textendash170 $\mu $m region, where instrumental limitations may have impacted the results, the observed $T_\text{e}(n_\text{e})$ relationship follows the adiabatic expansion scaling, $T_\text{e}\propto n_\text{e}^{\gamma-1}$, with  $\gamma=5/3$, consistent with expansion of a monatomic ideal gas. This agreement suggests that radiative losses do not dominate the energy balance at distances beyond 170 $\mu$m. 

One should also consider the possibility that the probe laser itself contributes to plasma heating. The energy of the probe laser pulse is absorbed through inverse Bremsstrahlung and transferred into the thermal electron energy. The maximum increase in electron temperature is then given by $\Delta T_{e,\max} = \frac{2}{3} \frac{\kappa_{\text{IB}} E_{\text{laser}}}{k_\text{B} \pi n_\text{e} r_0^2}$, where $\kappa_{\text{IB}}$ is the inverse Bremsstrahlung absorption coefficient for a fully ionized plasma, $k_\text{B}$ is the Boltzmann constant, $E_{\text{laser}}$ is the probe laser energy (20 mJ), $n_{\text{e}}$ is the plasma density, and $r_{\text{0}}$ is the laser spot radius (20 $\mu$m). The full $\kappa_{\text{IB}}$ formula is given elsewhere.\cite{lochte-holtgreven1995, bak2024} 

The effect of plasma heating in dense low temperature plasmas, such as laser-produced plasmas, can be very significant. However, a more accurate estimation has to account for the thermal diffusion outside of the laser heated volume.\cite{sato2018, lochte-holtgreven1995, pan2023b} During the laser pulse duration, $\Delta t$, the heat expands over the distance $\Delta r=\sqrt{\chi \Delta t}$, where $\chi=\frac{\kappa}{\frac{5}{2}n_\text{e}k_\text{B}}$ is the plasma thermal diffusivity, and $\kappa$ is the Spitzer thermal conductivity.\cite{sato2018, lochte-holtgreven1995, spitzer1962} Consequently, the laser energy is effectively distributed over the linear dimension of $r=r_0 + \Delta r$, and the value of $\Delta T_{e,\max}$ found above should by multiplied by a factor of $(\frac{r_0}{r_0 + \Delta r})^2 $. Here, for the experimentally measured values of $n_\text{e}$ and $T_\text{e}$, and assuming the average charge state $Z$ between 8\textendash12, the thermal diffusion length was calculated to be on the order of 150\textendash300 $\mu$m. Then, the plasma heating effect, $\frac {\Delta T_{e,\max}} {T_\text{e}}$, was below 13\% for all conditions explored in this work.

As for the electron density, for high-power EUV generation, $n_\text{e}$ must be high enough for efficient laser absorption but low enough to minimize opacity. For 1064 nm laser-generated tin plasmas at optimal $T_\text{e}$ between 25\textendash50 eV, the preferred $n_\text{e}$ range is considered to be between $10^{18}\text{--}10^{19}$ cm$^3$. \cite{versolato2019, bakshi2023, tomita2017} Here, at distances between 120\textendash270 $\mu$m from the target, the electron density was observed to decrease from $ 10^{18}$ to $ 10^{17} \text{ cm}^{-3}$ (Figure \ref{boxplot_coherent}b). While the region closer to the target was inaccessible for the current detection setup, the electron density is expected to monotonically increase toward the target surface. \cite{harilal2011, polek2025, tao2005} Therefore, although the measurements between 120\textendash150 $\mu $m show a potential plateau near $n_\text{e} \approx 1.8 \times 10^{18} \text{ cm}^{-3}$, we attribute this to the experimental limitations described above rather than to an actual plasma behavior. This conclusion was further confirmed by laser interferometry measurements that showed the electron density monotonically increasing toward the target (Section \ref{laserinterferometry}).

Next, we analyze the IAW feature of a Thomson scattering spectrum, which arises from the collective motion of ions. At conditions typical to tin EUV plasmas, the IAW feature appears as a doublet with the peaks separated by 50\textendash150 pm and the central frequency shifted from the laser line due to the plasma fluid motion.\cite{tomita2015, tomita2017, pan2023a} The spectral shift is directly correlated with the plasma fluid velocity as $v_k = \frac{c\Delta\lambda}{\lambda_\text{laser} 2\sin(\theta/2)}$, where $v_k$ is the velocity component along the scattering vector $\mathbf{k}$ (Figure \ref{ThomsonSetup}), $c$ is the speed of light, $\Delta\lambda$ is the shift relative to the laser line $\lambda_\text{laser}$, and $\theta$ is the scattering angle. \cite{sheffield2011, sande2002} Here, as shown in Figure \ref{Thomsonexample}, the IAW feature was under-resolved due to the finite spectral resolution. Therefore, we determined the spectral shift based on the peak position of a Gaussian profile fitted to the feature. The region rejected by the VBG notch filters was excluded from the fit.

\begin{figure}
    \centering
    \includegraphics[width=0.9\linewidth]{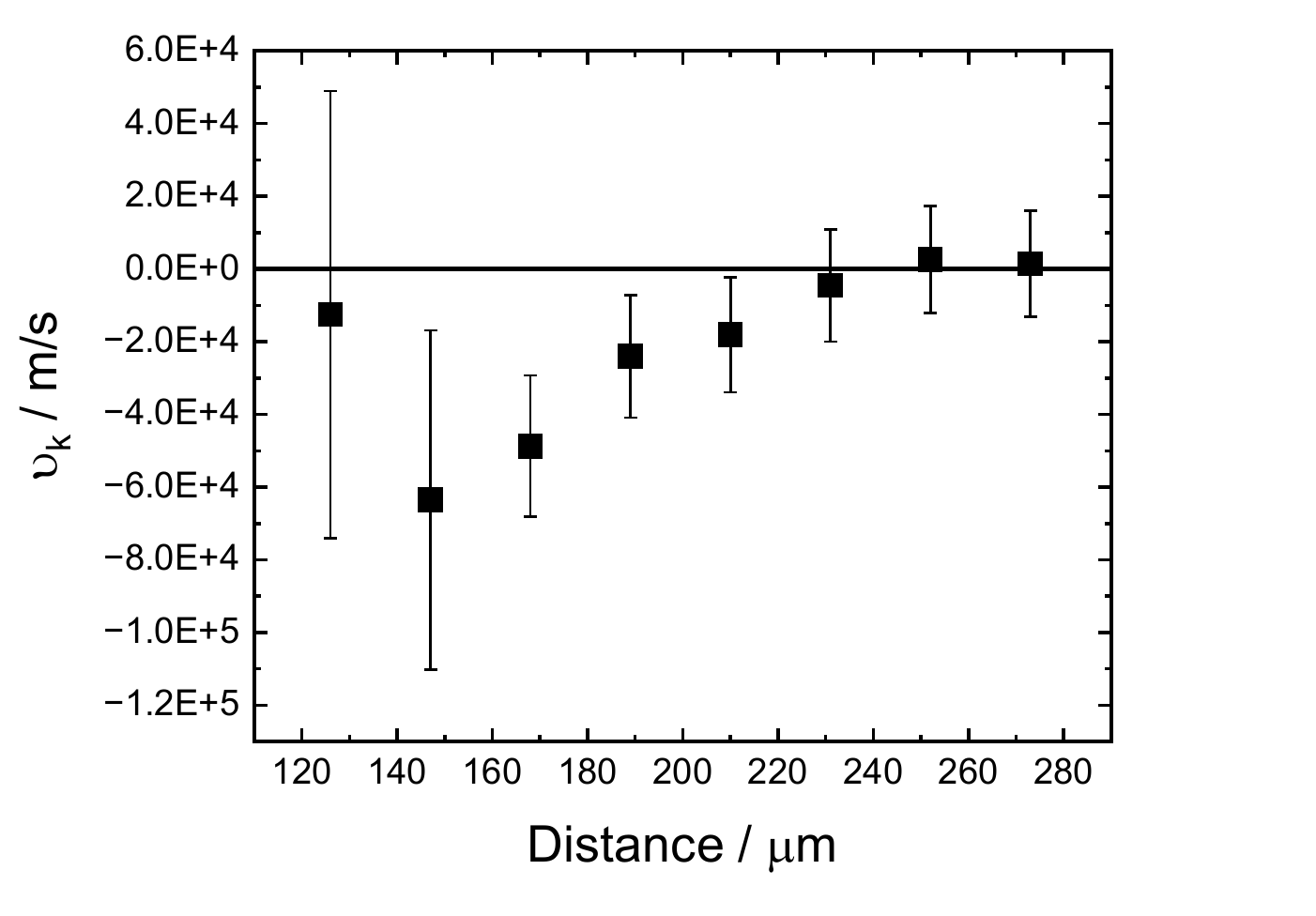}
    \caption{Plasma fluid velocity component ($v_k$) calculated from the shift of the Thomson scattering spectrum. The error bars denote $\pm2\sigma$ uncertainty.}
    \label{velocity}
\end{figure}

Figure \ref{velocity} shows the values of $v_k$ calculated from the corresponding spectral shifts versus the distance from the target. These experiments were performed twice using different gratings, 1200 g/mm and 2400 g/mm, with corresponding wavelength dispersions of 0.03 nm/pixel and 0.01 nm/pixel. The two datasets yielded similar results and were combined into a single plot. The error bars denote $\pm2\sigma$ uncertainty, incorporating the uncertainty of the Gaussian fit of the IAW feature and potential VBG filters misalignment, which affects their spectral transmission.

Measured velocity values range from $10^4\text{--}10^5$ m/s near the target to approximately zero at larger distances. The absolute magnitudes are consistent with literature for tin laser-produced plasmas. \cite{tomita2017} However, laser-produced plasmas are expected to continuously accelerate along the expansion axis driven by steep pressure gradients,\cite{sheffield2011, doggett2011} while we observed a reduction of $v_k$ with distance. This may stem from geometric factors, such as plume tilt or lateral misalignment between the probe and drive beams, which would cause the primary velocity vector to deviate from $\mathbf{k}$, resulting in a diminished $v_k$ projection. Moreover, the study by Tomita et al. revealed that the internal plasma dynamics may be more complex than a simple outward expansion, involving inward flows from the periphery toward the central region. \cite{tomita2023} In the next set of experiments, we intend to perform angle-resolved Thomson scattering measurements to distinguish between the geometric misalignment factors and intrinsic complex plasma flows.

\subsection{Laser Interferometry} \label{laserinterferometry}
\begin{figure} [b]
    \centering
    \includegraphics[width=1\linewidth]{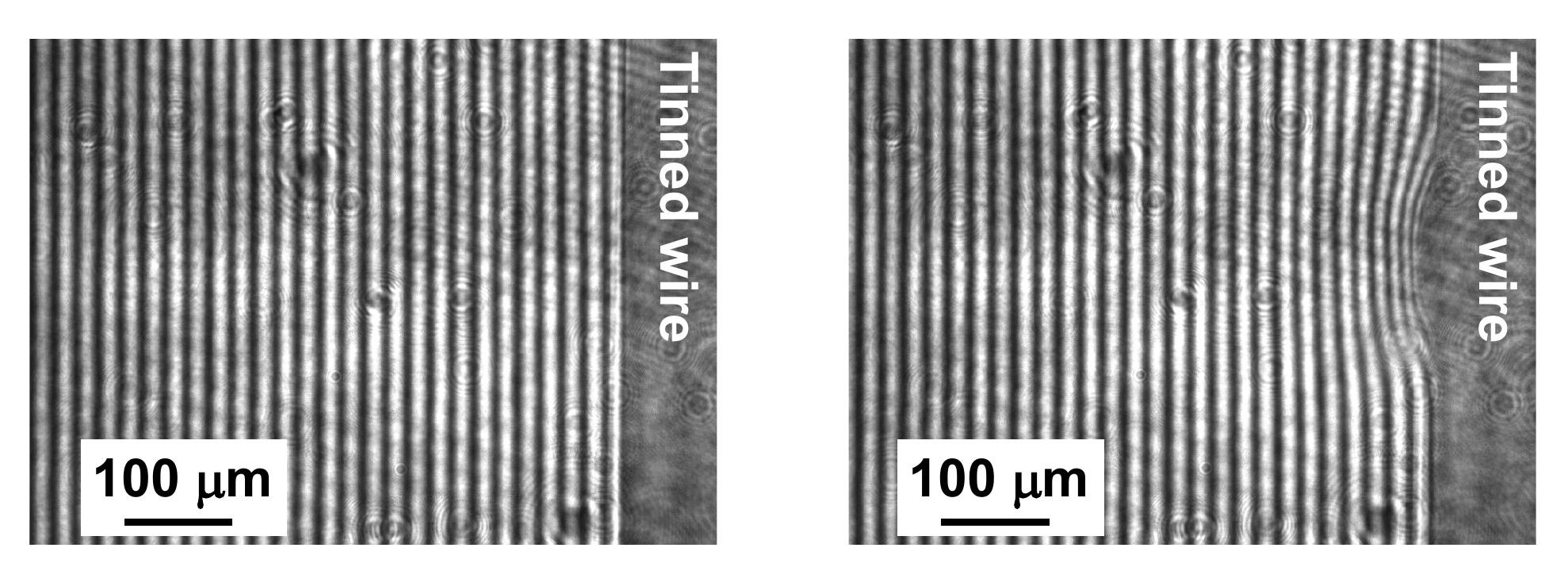}
    \caption{Examples of the reference (left) and plasma-distorted interferograms (right).}
    \label{InterferometryExample}
\end{figure}

\begin{figure}
    \centering
    \includegraphics[width=0.9\linewidth]{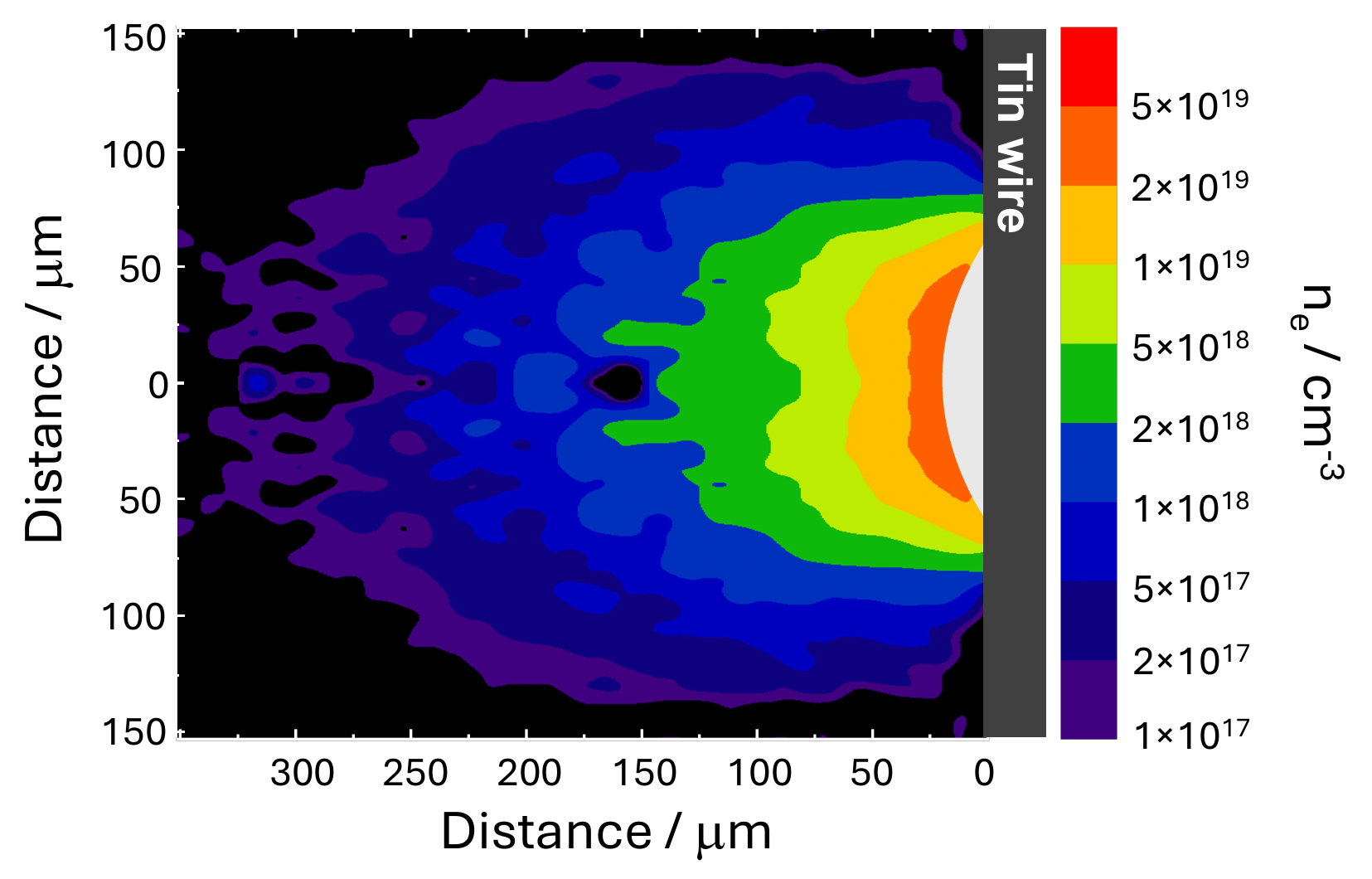}
    \caption{Example of a reconstructed 2D electron density map. Grayed out region was not accessible due to laser beam refraction and opacity effects.}
    \label{2Ddensity}
\end{figure}

To obtain electron densities closer to the target and to cross-validate Thomson scattering analysis, we conducted laser interferometry measurements. Figure \ref{InterferometryExample} shows representative reference and plasma-distorted interferograms. Figure \ref{2Ddensity} demonstrates the resulting 2D electron density maps obtained via inverse Abel transform, assuming cylindrical symmetry of the plasma plume.

\begin{figure}
    \centering
    \includegraphics[width=0.8\linewidth]{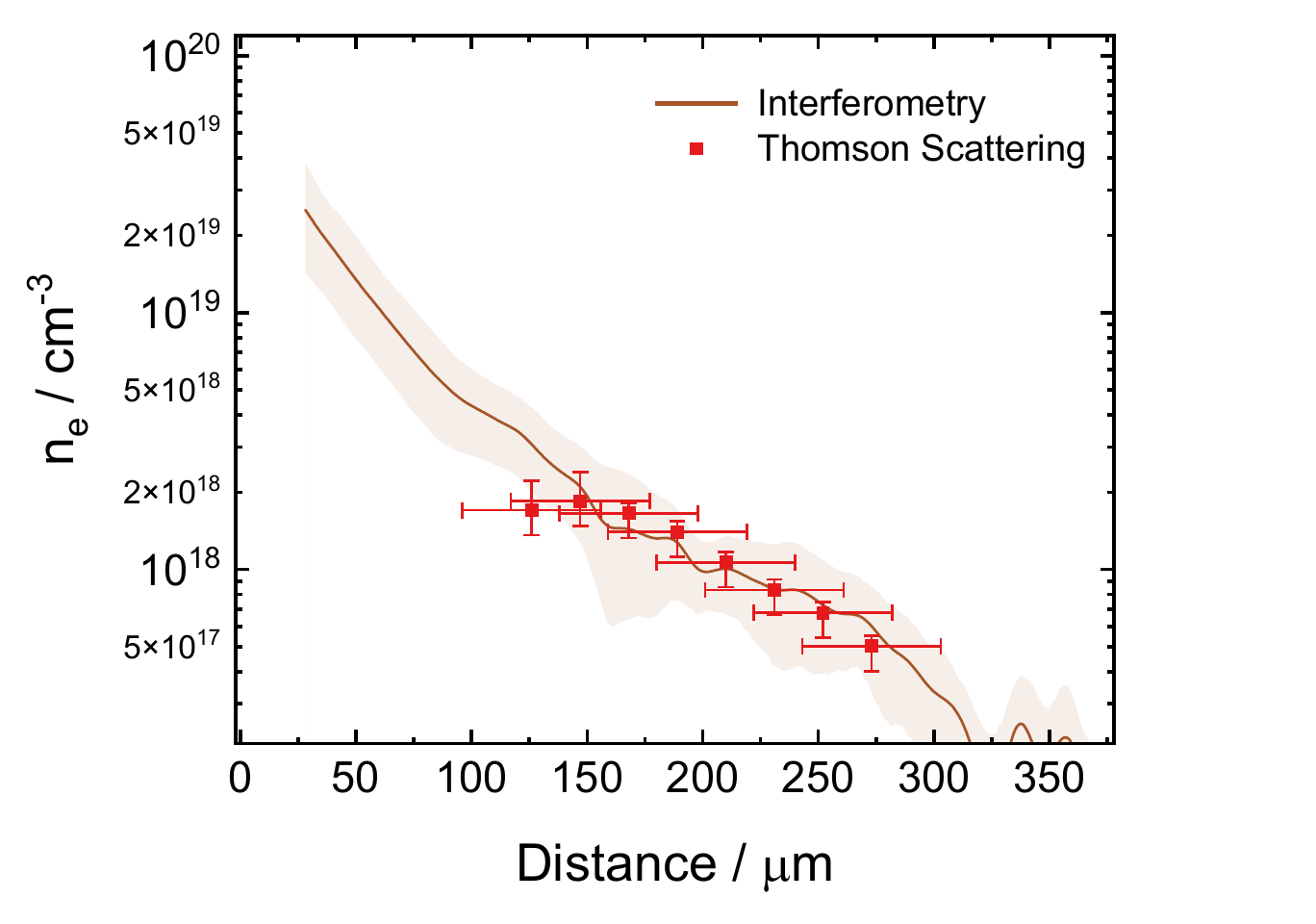}
    \caption{Electron density ($n_\text{e}$) spatial profiles measured by laser interferometry and coherent Thomson scattering. The shaded area corresponds to the 10\textendash90th percentile. The sensitivity limit of laser interferometry was $1 \times 10^{17} \text{ cm}^{-3}$.}
    \label{interf-vs-thomson}
\end{figure}

Figure \ref{interf-vs-thomson} compares the spatial profiles of electron densities measured with laser interferometry (along the symmetry axis) and coherent Thomson scattering. The shaded region indicates the 10\textendash90th percentile of the interferometry data that were collected over 150 laser shots. In the overlapping measurement region (120\textendash270 $\mu$m from the target), electron densities derived from laser interferometry and coherent Thomson scattering agree within experimental uncertainty. This cross-validation strengthens confidence in both diagnostic techniques and confirms the reliability of the $n_\text{e}$ measurements.

The maximum measurable density was $3 \times 10^{19}$ cm$^{-3}$, located approximately 20\textendash40 $\mu$m from the target. Densities closer to the target could not be accessed with the current experimental arrangement. Although the critical density for a 532 nm probe is $3.9 \times 10^{21}$ cm$^{-3}$, the practical upper limit is constrained by refraction and opacity. This region, indicated by the shaded gray area in Figure \ref{2Ddensity}, corresponds directly to the fringe-free zone observed in Figure \ref{InterferometryExample}b. Similar upper limit densities, measurable with 532 nm light, were obtained in other studies on plasmas of similar sizes. \cite{harilal2011, park2016, polek2025, tao2005} The lower limit (sensitivity) was determined by the minimum detectable fringe shift and, assuming a 500 $\mu$m thick plasma column, was found to be between $1\times 10^{17}$ to $3\times 10^{17}$cm$^{-3}$, depending on the quality of a particular interferogram.

\begin{figure}
    \centering
    \includegraphics[width=0.8\linewidth]{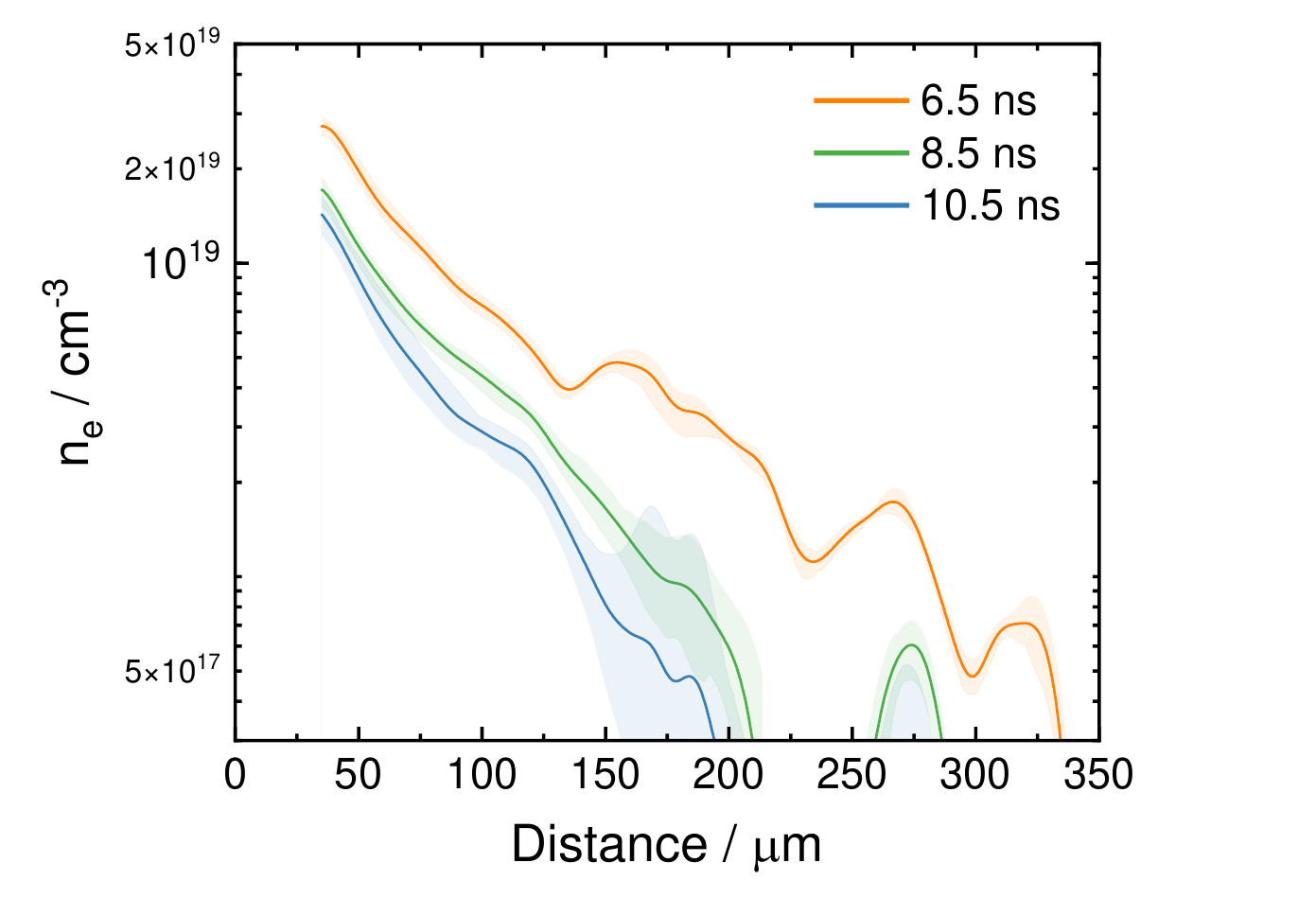}
    \caption{Axial electron densities ($n_\text{e}$) measured at different delays relative to the peak of the ablation pulse. The shaded regions indicate the 10\textendash90th percentile for each curve. The sensitivity limit was $3 \times 10^{17} \text{ cm}^{-3}$ for all curves.}
    \label{interf-time}
\end{figure}

We also used laser interferometry to characterize the temporal evolution of the electron density. Different delay times between the probe and ablation laser pulses were achieved by adjusting the optical path length of the probe laser. Figure \ref{interf-time} shows electron densities measured along the plume symmetry axis.
The peak electron densities decrease over time from $3 \times 10^{19} \text{ cm}^{-3}$ at 6.5 ns to $1.5 \times 10^{19} \text{ cm}^{-3}$ at 10.5 ns. The apparent non-monotonic behavior, especially at longer distances, is attributed to shot-to-shot variations in produced plasmas and to a measurement sensitivity limit of $3 \times 10^{17} \text{ cm}^{-3}$. Nevertheless, it can be clearly seen that the plasma density decreases with time indicating the expansion of the plasma plume. 

\section{Summary}
We comprehensively characterized laser-produced tin (Sn) plasmas using a multi-diagnostic approach consisting of coherent Thomson scattering, laser interferometry, and EUV emission spectroscopy. The diagnostics were implemented on a new experimental platform, "SparkLight", in which tin plasmas are generated via laser irradiation (1064 nm, 10 ns, up to $5.7\times 10^{10}$ W/cm$^2$) of a continuously moving tin-coated wire. Coherent Thomson scattering yielded electron temperatures of 9\textendash15 eV and densities of $5\times 10^{17}$ to $2\times 10^{18}$ cm$^{-3}$ at distances of 120\textendash270 $\mu$m from the target. The obtained densities were cross-validated against laser interferometry measurements, showing excellent agreement. 

Complementary EUV emission spectroscopy was used to assess spectral purity, estimate tin ion charge states, and map the spatial distribution of the EUV emission. Here, for the laser intensity of $ 10^{10}$ W/cm$^2$, the bulk of EUV emission originated within 150 $\mu$m of the target. The intensity peaked between 13.6\textendash13.8 nm, i.e., only partially overlapping with the in-band region, implying suboptimal plasma conditions. 

Future upgrades to "SparkLight" will include increased drive and probe laser energies and expanded detection spectral range to enable characterization of the near-target region where the bulk of EUV emission originates.

\begin{acknowledgments}
The research described in this paper was conducted under the Laboratory Directed Research and Development (LDRD) Program at Princeton Plasma Physics Laboratory, a national laboratory operated by Princeton University for the U.S. Department of Energy under Prime Contract No. DE-AC02-09CH11466. We thank Dr. Kirill Lezhnin (PPPL), Dr. Margarita Rivers (PPPL), and Prof. Dustin Froula (University of Rochester) for fruitful discussions.
\end{acknowledgments}

\section*{Data Availability Statement}
The data that supports the findings of this study are openly available at https://doi.org/10.34770/ajd8-by34.

\bibliography{references}

\end{document}